\documentclass[a4paper,12pt]{amsart}
\usepackage{natbib}
\usepackage[T1]{fontenc}
\usepackage{amsmath}
\usepackage[obeyspaces]{url}
\usepackage{soul}

\title{On a theory of electricity and gravitation}

\thanks{Originally published in German as ''Zur Elektricit{\"a}ts- und Gravitationstheorie'', \textit{\"Oversigt af Finska Vetenskaps-Societetens F\"orhandlingar} (Helsingfors), Bd. LVII. 1914-1915. Afd. A. N:o 4, p. 1--15. Translated by Frank Borg, present address: University of Jyv\"askyl\"a, Chydenius Institute, POB 567, 67101-Karleby, Finland; email: \url{borgbros@netti.fi}. A scanned pdf-version of the original paper along with the two other papers that Nordstr\"om published on the topic can be found at this url: \url{www.netti.fi/~borgbros/nordstrom}. For some brief historical sketches see Ch.\ Cronstr{\"o}m, ''Gunnar Nordstr{\"o}m (1881--1923)'' p.\ 3--9; S.\ Deser, ''The many dimensions of dimension'', p.\ 65--74; F.\ Ravndal, ''Scalar gravitation and extra dimensions'', p.\ 151--164: in Ch.\ Cronstr{\"o}m and C.\ Montonen (eds), \textit{Proceedings of the Gunnar  Nordstr{\"o}m Symposium on Theoretical Physics}, August 27--30, 2003, Helsinki. Commentationes Physico-Mathematicae 166/2004, The Finnish Society of Sciences and Letters.} 
\author{Gunnar Nordstr\"om}

\begin{document}

\begin{abstract}
This is the second paper by Nordstr\"om on his five dimensional theory. It aims at generalizing the relations $\mathbf{D} = \epsilon \mathbf{E}$, $\mathbf{B} = \mu \mathbf{H}$ of electromagnetism to the five dimensional gravito-electric theory and to show that this is consistent with the principle of equivalence. Some steps are taken toward providing a foundation for a theory of matter. 
Nordstr\"om hopes, as stated in the conclusion, to show ''that the five dimensional perspective, while still having only a formal content, may however provide footholds for further developments of the theory of gravitation, which otherwise would have been hard to find''. -- F.B.

\end{abstract}

\maketitle

In a previous communication\footnote{\so{G. Nordstr{\"o}m}, Phys.\ Zeitschr.\ \textbf{15}, p.\ 375, 1914. [The page reference should be 504. -- F.B.]} I have demonstrated how one can unite the equations of gravitation and the electromagnetic fields in a symmetrical fashion, if one considers the four dimensional spacetime-world to be a surface situated in a five dimensional world-extension.\footnote{I borrow the expression world-extension [Welterweiterung] from a letter by prof.\ \so{A. Sommerfeld}.} The cited communication concerned itself mainly with the field in an empty space and therefore the right hand side of the equations (I) was not scrutinized. Neither was any distinction made between field strength and field intensity [Erregung], since these concepts coincide in empty space. In the present communication these restrictions will be lifted, and it will be shown how to write the right hand side of the main equations (I) such as to correspond to the gravitational theory previously developed by me. 

In the general case the field is characterized by two ten-vectors $\mathfrak{f}$ and $\mathfrak{F}$, which coincide in the empty space, and whose relation at points of matter are given by certain ''supplementary conditions''. The two main systems of equations\footnote{The main 15 equations are naturally not independent. Among the equations (I) 4 are independent, among the equations (II) 6 are independent.} (I) and (II) become, when instead of $\frac{1}{c}\mathfrak{f}$ in the earlier communication we now use $\mathfrak{s}$,

{
\allowdisplaybreaks
\begin{align}
\label{EQ:I}
\nonumber
\frac{\partial \mathfrak{f}_{xy}}{\partial y} +
\frac{\partial \mathfrak{f}_{xz}}{\partial z} +
\frac{\partial \mathfrak{f}_{xu}}{\partial u} +
\frac{\partial \mathfrak{f}_{xw}}{\partial w} = \mathfrak{s}_x ,\\
\nonumber
\frac{\partial \mathfrak{f}_{yx}}{\partial x} +
\frac{\partial \mathfrak{f}_{yz}}{\partial z} +
\frac{\partial \mathfrak{f}_{yu}}{\partial u} +
\frac{\partial \mathfrak{f}_{yw}}{\partial w} = \mathfrak{s}_y ,\\
\tag{I}
\frac{\partial \mathfrak{f}_{zx}}{\partial x} +
\frac{\partial \mathfrak{f}_{zy}}{\partial y} +
\frac{\partial \mathfrak{f}_{zu}}{\partial u} +
\frac{\partial \mathfrak{f}_{zw}}{\partial w} = \mathfrak{s}_z ,\\
\nonumber
\frac{\partial \mathfrak{f}_{ux}}{\partial x} +
\frac{\partial \mathfrak{f}_{uy}}{\partial y} +
\frac{\partial \mathfrak{f}_{uz}}{\partial z} +
\frac{\partial \mathfrak{f}_{uw}}{\partial w} = \mathfrak{s}_u ,\\
\nonumber
\frac{\partial \mathfrak{f}_{wx}}{\partial x} +
\frac{\partial \mathfrak{f}_{wy}}{\partial y} +
\frac{\partial \mathfrak{f}_{wz}}{\partial z} +
\frac{\partial \mathfrak{f}_{wu}}{\partial u} = \mathfrak{s}_w ,
\end{align}

\begin{align}
\label{EQ:II}
\nonumber
\frac{\partial \mathfrak{F}_{yz}}{\partial x} +
\frac{\partial \mathfrak{F}_{zx}}{\partial y} +
\frac{\partial \mathfrak{F}_{xy}}{\partial z} = 0 ,\\
\nonumber
\frac{\partial \mathfrak{F}_{zu}}{\partial y} +
\frac{\partial \mathfrak{F}_{uy}}{\partial z} +
\frac{\partial \mathfrak{F}_{yz}}{\partial u} = 0 ,\\
\nonumber
\frac{\partial \mathfrak{F}_{xu}}{\partial z} +
\frac{\partial \mathfrak{F}_{uz}}{\partial x} +
\frac{\partial \mathfrak{F}_{zx}}{\partial u} = 0 ,\\
\nonumber
\frac{\partial \mathfrak{F}_{yu}}{\partial x} +
\frac{\partial \mathfrak{F}_{ux}}{\partial y} +
\frac{\partial \mathfrak{F}_{xy}}{\partial u} = 0 ,\\
\nonumber
\frac{\partial \mathfrak{F}_{zw}}{\partial y} +
\frac{\partial \mathfrak{F}_{wy}}{\partial z} +
\frac{\partial \mathfrak{F}_{yz}}{\partial w} = 0 ,\\
\tag{II}
\frac{\partial \mathfrak{F}_{xw}}{\partial z} +
\frac{\partial \mathfrak{F}_{wz}}{\partial x} +
\frac{\partial \mathfrak{F}_{zx}}{\partial w} = 0 ,\\
\nonumber
\frac{\partial \mathfrak{F}_{yw}}{\partial x} +
\frac{\partial \mathfrak{F}_{wx}}{\partial y} +
\frac{\partial \mathfrak{F}_{xy}}{\partial w} = 0 ,\\
\nonumber
\frac{\partial \mathfrak{F}_{uw}}{\partial x} +
\frac{\partial \mathfrak{F}_{wx}}{\partial u} +
\frac{\partial \mathfrak{F}_{xu}}{\partial w} = 0 ,\\
\nonumber
\frac{\partial \mathfrak{F}_{uw}}{\partial y} +
\frac{\partial \mathfrak{F}_{wy}}{\partial u} +
\frac{\partial \mathfrak{F}_{yu}}{\partial w} = 0 ,\\
\nonumber
\frac{\partial \mathfrak{F}_{uw}}{\partial z} +
\frac{\partial \mathfrak{F}_{wz}}{\partial u} +
\frac{\partial \mathfrak{F}_{zu}}{\partial w} = 0 .
\end{align}
}

The ten-vector $\mathfrak{F}$ may be derived from a five-potential $\varPhi$ according to the 10 equations

\begin{equation}
\label{EQ:1}
\mathfrak{F}_{mn} = \frac{\partial \varPhi_n}{\partial m} -
\frac{\partial \varPhi_m}{\partial n} .
\end{equation}

Next we turn to the right hand side of the equations (\ref{EQ:I}); that is, to components of the five-current. According to the gravitational theory that we will adhere to here\footnote{\so{G. Nordstr{\"o}m}, Ann.\ d.\ Phys.\ \textbf{42}, p.\ 533, 1913.}, the mass-density is determined by the matter-elasticity tensor $\mathsf{T}$. Due to the five dimensional formulation the tensor is extended via the components

\begin{equation*}
\mathsf{T}_{wx}, \mathsf{T}_{wy}, \mathsf{T}_{wz}, \mathsf{T}_{wu}, \mathsf{T}_{ww}
\end{equation*} 

to a five dimensional tensor. The five-current $\mathfrak{s}$ is to be expressed in terms of the extended tensor $\mathsf{T}$ and the five-potential $\varPhi$. However, it turns out that it is advantageous to introduce a tensor $\mathsf{S}$ in place of $\mathsf{T}$, whose off-diagonal terms coincide with those of $\mathsf{T}$ as

\begin{equation}
\label{EQ:2}
\left\{
\begin{aligned}
\mathsf{S}_{xx} &= \mathsf{T}_{xx} - 
\left(
\mathsf{T}_{xx} + \mathsf{T}_{yy} + \mathsf{T}_{zz} + \mathsf{T}_{uu} + \mathsf{T}_{ww}
\right)
\\
&= -
\left(
\mathsf{T}_{yy} + \mathsf{T}_{zz} + \mathsf{T}_{uu} + \mathsf{T}_{ww}
\right)
\\
&\mbox{etc.}
\\
\mathsf{S}_{xy} &= \mathsf{T}_{xy} \quad \mbox{etc.} 
\end{aligned}
\right.
\end{equation}

We will find that we arrive at the gravitational theory that I have developed if we write the five-current as:

\begin{equation}
\label{EQ:3}
\left\{
\begin{aligned}
\mathfrak{s}_x &= -
\frac{\mathsf{S}_{xx} \mathfrak{n}_x + \mathsf{S}_{xy} \mathfrak{n}_y + \mathsf{S}_{xy} \mathfrak{n}_y + \mathsf{S}_{xu} \mathfrak{n}_u + \mathsf{S}_{xw} \mathfrak{n}_w}{\varPhi_x \mathfrak{n}_x + \varPhi_y \mathfrak{n}_y + \varPhi_z \mathfrak{n}_z + \varPhi_u \mathfrak{n}_u + \varPhi_w \mathfrak{n}_w},
\\
&\dots
\\
\mathfrak{s}_w &= -
\frac{\mathsf{S}_{wx} \mathfrak{n}_x + \mathsf{S}_{wy} \mathfrak{n}_y + \mathsf{S}_{wy} \mathfrak{n}_y + \mathsf{S}_{wu} \mathfrak{n}_u + \mathsf{S}_{ww} \mathfrak{n}_w}{\varPhi_x \mathfrak{n}_x + \varPhi_y \mathfrak{n}_y + \varPhi_z \mathfrak{n}_z + \varPhi_u \mathfrak{n}_u + \varPhi_w \mathfrak{n}_w},
\end{aligned}
\right.
\end{equation}

where $\mathfrak{n}$ is a five-vector to be discussed. In order to arrive at the said gravitational theory we must take $\mathfrak{n}$ to be a vector which is normal to the four dimensional world-surface, whose placement it thus determines. Furthermore we have to assume that the world-surface is planar, and that the derivatives of all the field quantities in the direction of $\mathfrak{n}$ are equal to zero. In order to see, that these assumptions lead to the desired goal, we choose the direction $\mathfrak{n}$ to be along the $w$-axis. The last equation (\ref{EQ:3}) yields thus

\begin{equation}
\label{EQ:3a}
\tag{3 a}
\mathfrak{s}_w = - \frac{\mathsf{S}_{ww}}{\varPhi_w} = 
\frac{1}{\varPhi_w} 
\left(
\mathsf{T}_{xx} + \mathsf{T}_{yy} + \mathsf{T}_{zz} + \mathsf{T}_{uu}
\right), 
\end{equation}

whence $- \mathfrak{s}_w$ is equal to the rest-density $g \nu$ of the gravitating mass\footnote{\so{G. Nordstr{\"o}m}, Ann.\ d.\ Phys.\ \textbf{42}, p.\ 537, 1913.}. The equations (\ref{EQ:I}) and (\ref{EQ:II}) give now, when we set $\mathfrak{F}_{wx}$ = $\mathfrak{f}_{wx}$ etc, immediately the fundamental equations of my theory of gravitation.

We have seen, that the component of $\mathfrak{s}$ normal to the world-surface gives the gravitating mass. For an arbitrary reference system one has

\begin{equation}
\label{EQ:4}
g  \nu = - \frac{1}{\sqrt{\mathfrak{n}^2}}
\left(
\mathfrak{n}_x \mathfrak{s}_x +
\mathfrak{n}_y \mathfrak{s}_y +
\mathfrak{n}_z \mathfrak{s}_z +
\mathfrak{n}_u \mathfrak{s}_u +
\mathfrak{n}_w \mathfrak{s}_w 
\right),
\end{equation}

where

\begin{equation}
\label{EQ:5}
\mathfrak{n}^2 = 
\mathfrak{n}_x^2 +
\mathfrak{n}_y^2 +
\mathfrak{n}_z^2 +
\mathfrak{n}_u^2 +
\mathfrak{n}_w^2 .
\end{equation}

The remaining components -- which thus lies in the world-surface -- of $\mathfrak{s}$ yield the electrical four-current. Its components are given by

\begin{equation}
\label{EQ:6}
\mathfrak{s}_x^e = \mathfrak{s}_x - \frac{\mathfrak{n}_x}{\mathfrak{n}^2}
\left(
\mathfrak{n}_x \mathfrak{s}_x +
\mathfrak{n}_y \mathfrak{s}_y +
\mathfrak{n}_z \mathfrak{s}_z +
\mathfrak{n}_u \mathfrak{s}_u +
\mathfrak{n}_w \mathfrak{s}_w 
\right)
\end{equation}

etc. When we \so{exclude the conduction current}, then the absolute magnitude of $\mathfrak{s}_x^e$ multiplied by $-i$ gives the electric rest density $\varrho_0$, 

\begin{equation}
\label{EQ:7}
\varrho_0 = - i \,
\sqrt{ 
{\mathfrak{s}_x^e}^2 +
{\mathfrak{s}_y^e}^2 +
{\mathfrak{s}_z^e}^2 +
{\mathfrak{s}_u^e}^2 +
{\mathfrak{s}_w^e}^2  
}. 
\end{equation} 

Furthermore, $\mathfrak{s}^e$ determines the motion vector (the four-velocity) $\mathfrak{B}$ according to

\begin{equation}
\label{EQ:8}
\frac{1}{c} \mathfrak{B}_x = \frac{\mathfrak{s}_x^e}{\varrho_0},
\dots \dots 
\frac{1}{c} \mathfrak{B}_w = \frac{\mathfrak{s}_w^e}{\varrho_0}.
\end{equation}

$c$ is the velocity of light.

In particular, if one chooses the $w$-direction to be along the normal of the world-surface, then one obtains

\begin{align}
\nonumber
\mathfrak{s}_w^e = 0, \; \mathfrak{B}_w = 0,
\\
\label{EQ:7a}
\tag{7 a}
\varrho_0 = - \frac{i}{\varPhi_w} 
\sqrt{\mathsf{S}_{xw}^2 + \mathsf{S}_{yw}^2 + \mathsf{S}_{zw}^2 +\mathsf{S}_{uw}^2},
\\
\label{EQ:8a}
\tag{8 a}
\mathfrak{s}_w^e = \frac{1}{c} \mathfrak{B} \varrho_0 = - \frac{\mathsf{S}_{xw}}{\varPhi_w}
\\
\nonumber
\mbox{etc}.
\end{align}

We will now move to the question related to the difference between the two ten-vectors $\mathfrak{f}$ and $\mathfrak{F}$. First we have to discuss a few algebraic vector operation. 

Given a five-vector ($\mathfrak{A}$) and ten-vector ($\mathfrak{B}$) then we can form, through multiplication, both a five-vector and ten-vector. The five-vector has the components

\begin{equation}
\label{EQ:9}
\mathfrak{C}_x = 
\mathfrak{A}_y \mathfrak{B}_{xy} +
\mathfrak{A}_z \mathfrak{B}_{xz} +
\mathfrak{A}_u \mathfrak{B}_{xu} +
\mathfrak{A}_w \mathfrak{B}_{xw}.
\end{equation}

For the remaining components we have of course corresponding expressions. The vector $\mathfrak{C}$ is normal to $\mathfrak{A}$. The ten-vector formed from $\mathfrak{A}$ and $\mathfrak{B}$ has the $xy$-component,

\begin{equation}
\label{EQ:10}
\mathfrak{D}_{xy} = 
\mathfrak{A}_z \mathfrak{B}_{uw} +
\mathfrak{A}_u \mathfrak{B}_{wz} +
\mathfrak{A}_w \mathfrak{B}_{zu};
\end{equation}  

the remaining components are obtained by exchanging the indices. For instance, when constructing the $yu$-component

\begin{equation*}
\mathfrak{D}_{yu} = 
\mathfrak{A}_x \mathfrak{B}_{wz} +
\mathfrak{A}_w \mathfrak{B}_{zx} +
\mathfrak{A}_z \mathfrak{B}_{xw};
\end{equation*}  

one has to observe that the order $yuxwz$ (as well as $yuwzx$ and $yuzxw$) follows from  $xyzuw$ by an \so{even number} of permutations of the elements. The ten-vector obtained by the operation (\ref{EQ:10}) differs from the original one in similar way that an axial vector differs from a polar one in the three dimensional vector analysis.

By multiplication of two five-vectors one may produce a ten-vector and a scalar. The ten-vector has the components

\begin{equation}
\label{EQ:11}
\mathfrak{B}_x \mathfrak{D}_y - \mathfrak{B}_y \mathfrak{D}_x \quad \mbox{etc.};
\end{equation}

the scalar is

\begin{equation}
\label{EQ:12}
\mathfrak{B}_x \mathfrak{D}_x +
\mathfrak{B}_y \mathfrak{D}_y +
\mathfrak{B}_z \mathfrak{D}_z +
\mathfrak{B}_u \mathfrak{D}_u +
\mathfrak{B}_w \mathfrak{D}_w. 
\end{equation}

If one applies (\ref{EQ:9}) using the vectors $-\frac{1}{\sqrt{\mathfrak{n}^2}} \mathfrak{n}$ and $\mathfrak{f}$ (or $\mathfrak{F}$) then one obtains a five-vector which lies in the world-surface and which only relates to the gravitational field. This gravitational vector has the $x$-component

\begin{equation}
\label{EQ:13}
-\frac{1}{\sqrt{\mathfrak{n}^2}}
\left(
\mathfrak{n}_y \mathfrak{f}_{xy} +
\mathfrak{n}_z \mathfrak{f}_{xz} +
\mathfrak{n}_u \mathfrak{f}_{xu} +
\mathfrak{n}_w \mathfrak{f}_{xw} 
\right) ;
\end{equation}

if one takes $\mathfrak{n}$ to be in the $w$-direction, then the expression becomes equal to $\mathfrak{f}_{wx}$.

By repeating the operation (\ref{EQ:10}) twice using the vectors $\frac{1}{\sqrt{\mathfrak{n}^2}} \mathfrak{n}$ and $\mathfrak{f}$ one obtains a ten-vector, which only relates to the electromagnetic field, and whose components in the planes, which contain the $w$-direction, vanish.

When we consider the field in a material medium, we must, as has been pointed out, distinguish between the two ten-vectors $\mathfrak{f}$ and $\mathfrak{F}$. How should the supplementary conditions relating $\mathfrak{f}$ and $\mathfrak{F}$ be formulated? This formulation should naturally be an extension of the corresponding relations which \so{Minkowski} presented for the electromagnetic field\footnote{\so{H. Minkowski}, G\"ott.\ Nachr.\ 1908, p.\ 85 and 86 equations \{C\} and \{D\}.}. This though can be attained using various assumptions. 

Since, as shown above, it is possible to separate the electromagnetic and gravitational fields, one may formulate the supplementary conditions for these two parts of the total field independently of each other. Thus one may e.g.\ assume the following 5 equations for the gravitational field (of which only four are independent)

\begin{align}
\label{EQ:14}
\mathfrak{n}_y \mathfrak{f}_{xy} +
\mathfrak{n}_z \mathfrak{f}_{xz} +
\mathfrak{n}_u \mathfrak{f}_{xu} +
\mathfrak{n}_w \mathfrak{f}_{xw} =
\\
\nonumber
\alpha \left(
\mathfrak{n}_y \mathfrak{F}_{xy} +
\mathfrak{n}_z \mathfrak{F}_{xz} +
\mathfrak{n}_u \mathfrak{F}_{xu} +
\mathfrak{n}_w \mathfrak{F}_{xw} 
\right).  
\end{align}

$\alpha$ is hereby a quantity analogous to the dielectric constant and the magnetic permeability, which are equal to 1 for the vacuum. In the simplest case one takes $\alpha$ to be equal to 1 also for the material medium. When one assumes the condition (\ref{EQ:14}) for the gravitational field, then 
it is not that easy to express the \so{Minkowski} conditions in terms of the five dimensional equations, and we will refrain from doing it here.

One may instead of (\ref{EQ:14}) consider other conditions obtained by applying the operations (\ref{EQ:9}) and (\ref{EQ:10}) to the vectors $\mathfrak{B}$, $\mathfrak{f}$ and $\mathfrak{F}$. The conditions constitute two systems, one consisting of five, the other of ten equations. Due to space limits we write down only two equations for each system:
 
\begin{equation}
\label{EQ:15}
\left\{
\begin{aligned} 
&\mathfrak{B}_y \mathfrak{f}_{xy} +
\mathfrak{B}_z \mathfrak{f}_{xz} +
\mathfrak{B}_u \mathfrak{f}_{xu} +
\mathfrak{B}_w \mathfrak{f}_{xw}  =
\\ 
&\varepsilon
\left(
\mathfrak{B}_y \mathfrak{F}_{xy} +
\mathfrak{B}_z \mathfrak{F}_{xz} +
\mathfrak{B}_u \mathfrak{F}_{xu} +
\mathfrak{B}_w \mathfrak{F}_{xw} 
\right),
\\
&\dots \dots
\\
&\mathfrak{B}_y \mathfrak{f}_{wx} +
\mathfrak{B}_z \mathfrak{f}_{wy} +
\mathfrak{B}_u \mathfrak{f}_{wz} +
\mathfrak{B}_w \mathfrak{f}_{wu}  =
\\ 
&\varepsilon
\left(
\mathfrak{B}_y \mathfrak{F}_{wx} +
\mathfrak{B}_z \mathfrak{F}_{wy} +
\mathfrak{B}_u \mathfrak{F}_{wz} +
\mathfrak{B}_w \mathfrak{F}_{wu} 
\right),
\end{aligned}
\right.
\end{equation}

\begin{equation}
\label{EQ:16}
\left\{
\begin{aligned} 
&\mathfrak{B}_z \mathfrak{F}_{uw} +
\mathfrak{B}_u \mathfrak{F}_{wz} +
\mathfrak{B}_w \mathfrak{F}_{zu} 
 =
\mu
\left(
\mathfrak{B}_z \mathfrak{f}_{uw} +
\mathfrak{B}_u \mathfrak{f}_{wz} +
\mathfrak{B}_w \mathfrak{f}_{zu} 
\right),
\\ 
&\mathfrak{B}_y \mathfrak{F}_{zu} +
\mathfrak{B}_z \mathfrak{F}_{uy} +
\mathfrak{B}_u \mathfrak{F}_{yz} 
 =
\mu
\left(
\mathfrak{B}_y \mathfrak{f}_{zu} +
\mathfrak{B}_z \mathfrak{f}_{uy} +
\mathfrak{B}_u \mathfrak{f}_{yz} 
\right)
.
\end{aligned}
\right.
\end{equation}

According to these equations the supplementary conditions for the electromagnetic and gravitational fields will not -- contrary to the case with the equations (\ref{EQ:14}) -- be independent of each other. The dielectric constant and the magnetic permeability will also affect the gravitational field; though $\varepsilon$ will play a role only in the case of moving bodies and non-stationary fields.

It is to be shown that the equations (\ref{EQ:15}), (\ref{EQ:16}) contain the {\so{Minkow\-ski}} conditions, and this is easily achieved when the normal direction of the world-surface is chosen to be the same as the $w$-direction, by which $\mathfrak{B}_w$ = 0. It remains to be shown that the 15 equations (\ref{EQ:15}), (\ref{EQ:16}) contain no internal contradiction. To this end one further specifies the reference system such that $\mathfrak{B}_x$ = $\mathfrak{B}_y$ = $\mathfrak{B}_z$ = 0 (transformation to the rest system). By this the equations relate each component of $\mathfrak{f}$ to the corresponding component of $\mathfrak{F}$, and no internal inconsistency can arise since each component figures only once.

As has been brought forth, it is possible to assume different supplementary conditions for the gravitational field. One may ask, however, whether the principle of equivalence by \so{Einstein} may depend on these conditions. We will demonstrate, that this principle is generally valid whatever the way one chooses the conditions.

When we choose the normal of the world-surface to be in the $w$-direction we can write

\[
\mathfrak{f}_{wx} = \mathfrak{f}_{x} \; \mbox{etc.,} \; 
\mathfrak{F}_{wx} = \mathfrak{F}_{x} \; \mbox{etc.},
\]

and $\mathfrak{f}_{x}, \dots \mathfrak{f}_{u}$ $\mathfrak{F}_{x}, \dots \mathfrak{F}_{u}$ are the components of two four-vectors in the world-surface. All derivatives with respect to $w$ are assumed to be zero, and we will consider only the relations only as four dimensional. The gravitational equation becomes, when we write simply $\varPhi$ instead of $\varPhi_w$,

\begin{equation}
\label{EQ:17}
\frac{\partial \mathfrak{f}_x}{\partial x} + 
\frac{\partial \mathfrak{f}_y}{\partial y} +
\frac{\partial \mathfrak{f}_z}{\partial z} +
\frac{\partial \mathfrak{f}_u}{\partial u} =
-g(\varPhi)  \nu ,
\end{equation}

\begin{equation}
\label{EQ:18}
\left\{
\begin{aligned}
\frac{\partial \mathfrak{f}_y}{\partial x} =
\frac{\partial \mathfrak{f}_x}{\partial y}, \;
\frac{\partial \mathfrak{f}_z}{\partial y} =
\frac{\partial \mathfrak{f}_y}{\partial z}, \;
\frac{\partial \mathfrak{f}_x}{\partial z} =
\frac{\partial \mathfrak{f}_z}{\partial x}, 
\\
\frac{\partial \mathfrak{f}_u}{\partial x} =
\frac{\partial \mathfrak{f}_x}{\partial u}, \;
\frac{\partial \mathfrak{f}_u}{\partial y} =
\frac{\partial \mathfrak{f}_y}{\partial u}, \;
\frac{\partial \mathfrak{f}_u}{\partial z} =
\frac{\partial \mathfrak{f}_z}{\partial u}. 
\end{aligned}
\right.
\end{equation}

The vector $\mathfrak{F}$ is derived from a scalar gravitational potential $\varPhi$;

\begin{equation}
\label{EQ:19}
\mathfrak{F}_x = -\frac{\partial \varPhi}{\partial x}, \;
\mathfrak{F}_y = -\frac{\partial \varPhi}{\partial y}, \;
\mathfrak{F}_z = -\frac{\partial \varPhi}{\partial z}, \;
\mathfrak{F}_u = -\frac{\partial \varPhi}{\partial u}.
\end{equation}

The gravitational potential $\varPhi$ is the retarded potential of the \so{free} gravitating mass, whose density $\varphi$ is given by the equation

\begin{equation}
\label{EQ:20}
\frac{\partial \mathfrak{F}_x}{\partial x} +
\frac{\partial \mathfrak{F}_y}{\partial y} +
\frac{\partial \mathfrak{F}_z}{\partial z} +
\frac{\partial \mathfrak{F}_u}{\partial u} =
- \varphi .
\end{equation}

We have thus

\begin{equation}
\label{EQ:21}
\varPhi = - \frac{1}{4 \pi} \int \frac{dv}{r} \varPhi\left(t - \frac{r}{c}\right)
+ \varPhi_a ,
\end{equation}

where $\varPhi_a$ is the value of the potential at infinity. 

The ponderomotoric force $\mathfrak{K}$ generated by the gravitational field is expressed by a tensor $\mathsf{G}$:

\begin{equation}
\label{EQ:22}
\mathfrak{K}_x = -\left\{
\frac{\partial \mathsf{G}_{xx}}{\partial x} +
\frac{\partial \mathsf{G}_{xy}}{\partial y} +
\frac{\partial \mathsf{G}_{xz}}{\partial z} +
\frac{\partial \mathsf{G}_{xu}}{\partial u}
\right\} \; \mbox{etc}.
\end{equation}

For the diagonal components of the tensor we obtain\footnote{In the case of the combined electromagnetic and gravitational fields we have for the combined field-tensor
\begin{equation*}
\begin{split}
\mathsf{G}_{xx} = \frac{1}{2}
\{
\mathfrak{f}_{xy} \mathfrak{F}_{xy} +
\mathfrak{f}_{xz} \mathfrak{F}_{xz} +
\mathfrak{f}_{xu} \mathfrak{F}_{xu} +
\mathfrak{f}_{xw} \mathfrak{F}_{xw} -
\mathfrak{f}_{yz} \mathfrak{F}_{yz} -
\mathfrak{f}_{yu} \mathfrak{F}_{yu} -
\\
\mathfrak{f}_{zu} \mathfrak{F}_{zu} -
\mathfrak{f}_{yw} \mathfrak{F}_{yw} -
\mathfrak{f}_{zw} \mathfrak{F}_{zw} -
\mathfrak{f}_{uw} \mathfrak{F}_{uw} 
\} 
\quad \mbox{etc}.
\end{split}
\end{equation*}
}

\begin{equation}
\label{EQ:23}
\mathsf{G}_{xx} = \frac{1}{2}
\left\{
\mathfrak{f}_x \mathfrak{F}_x -
\mathfrak{f}_y \mathfrak{F}_y -
\mathfrak{f}_z \mathfrak{F}_z -
\mathfrak{f}_u \mathfrak{F}_u
\right\} \quad \mbox{etc}.
\end{equation} 

The off-diagonal components are of no consequence for the following considerations. 

As in Ann.\ d.\ Phys.\ 42, p.\ 535 we consider a ''complete, stationary system'' and choose the reference system such that the gravitational field becomes static. Since $\frac{\partial \varPhi}{\partial u} = 0$ one obtains for the diagonal sum of the gravitational tensor the expression

\[
\mathfrak{f}_x \frac{\partial \varPhi}{\partial x} +
\mathfrak{f}_y \frac{\partial \varPhi}{\partial y} +
\mathfrak{f}_z \frac{\partial \varPhi}{\partial z},
\]  

and when we proceed exactly as in Ann.\ d.\ Phys.\ 42, p.\ 536 we obtain according to the theorem by \so{Laue}

\begin{align*}
\int \left\{
-D + \mathfrak{f}_x \frac{\partial \varPhi}{\partial x} +
\mathfrak{f}_y \frac{\partial \varPhi}{\partial y} +
\mathfrak{f}_z \frac{\partial \varPhi}{\partial z}
\right\}dv =
\\
\int \left\{
\mathsf{T}_{uu} + \mathsf{G}_{uu} + \mathsf{L}_{uu}
\right\}dv = -E_0.
\end{align*}

-$D$ is the diagonal sum $\mathsf{T}_{xx} + \mathsf{T}_{yy} +\mathsf{T}_{zz} +\mathsf{T}_{uu}$ of the matter-elastic tensor $\mathsf{T}$, $E_0$ is the rest energy of the system.

Since in our case $\frac{\partial \mathfrak{f}_u}{\partial u} = 0$, we have

\begin{align*}
\frac{\partial}{\partial x} \varPhi \mathfrak{f}_x +
\frac{\partial}{\partial y} \varPhi \mathfrak{f}_y +
\frac{\partial}{\partial z} \varPhi \mathfrak{f}_z =
\\
-\varPhi g(\varPhi)  \nu +  
\mathfrak{f}_x \frac{\partial \varPhi}{\partial x} +
\mathfrak{f}_y \frac{\partial \varPhi}{\partial y} +
\mathfrak{f}_z \frac{\partial \varPhi}{\partial z}, 
\end{align*}

and obtain by integrating over the whole $xyz$-space and using \so{Gauss'} theorem

\begin{equation*}
\int \left\{
\mathfrak{f}_x \frac{\partial \varPhi}{\partial x} +
\mathfrak{f}_y \frac{\partial \varPhi}{\partial y} +
\mathfrak{f}_z \frac{\partial \varPhi}{\partial z}
\right\}dv =
\int \left(
\varPhi - \varPhi_a
\right) g(\varPhi)  \nu dv.
\end{equation*}

Our equations give now for the inertial mass $m = \frac{1}{c^2} E_0$ of the system the former expression

\begin{equation}
\label{EQ:24}
m = \frac{1}{c^2} \int 
\left\{
D - (\varPhi - \varPhi_a)g(\varPhi) \nu
\right\}dv.
\end{equation}

From this one proves exactly as in Ann.\ d.\ Phys.\ 42, p.\ 536 and 537 that the relation

\begin{equation}
\label{EQ:25}
g(\varPhi_a)m = \int g(\varPhi) \nu dv
\end{equation}

is valid when $\nu$ and $g(\varPhi)$ are defined by the equations

\begin{equation}
\label{EQ:26}
\nu = \frac{1}{c^2} D, \;
g(\varPhi) = \frac{c^2}{\varPhi},
\end{equation}

which according to equation (\ref{EQ:3a}) conform with our initial assumptions.

The equation (\ref{EQ:25}) expresses \so{Einstein}'s principle of equivalence when assuming that the gravitating mass 

\[
M_g = \int g(\varPhi) \nu dv
\]

determines for the gravitational action, which the system effects and by which it is affected. That this condition is satisfied in the present case requires some further elaboration. So then, in empty space $\mathfrak{f}$ and $\mathfrak{F}$ are identical. For the absolute magnitude of the field-vector at distant points we may find an expression, when we apply the integral equality

\[
\int \mathfrak{f}_n df = - \int g(\varPhi) \nu dv,
\]

derived from (\ref{EQ:17}), on a sphere, whose center point is occupied by the system and whose surface passes through the distant point of interest. Since the field-vector is radial on the surface of the sphere, the expression becomes:

\[
|\mathfrak{f}| = |\mathfrak{F}| = \frac{1}{4 \pi r^2} \int g(\varPhi) \nu dv.
\]

The gravitational field of caused by the system depends for large distances, as one can see, only on the gravitating mass of the system. That also the ponderomotoric force effected by an external, homogeneous field, does only depend on $M$, follows from the fact -- since the field is static -- that the force may be represented by fictional stresses inside of a surface which encloses the system and is very far away from it. Any torque acting on the system will however not be determined by $M_g$.

From what has been said, it follows from equation (\ref{EQ:25}) that the principle of equivalence by \so{Einstein} is valid in the theory developed, independently of what relation exists between $\mathfrak{f}$ and $\mathfrak{F}$.
By different choices of the supplementary conditions one arrives at different modifications of the gravitational theory. When the two vectors of the gravitational field are different, then the gravitational action is indeed dependent on the intervening medium. To what extent these relations can be investigated experimentally is a question which will not be dealt with here; the aim has only been to point to the possible modifications of the theory. 

In the present communication, as in my other communications on the topic of gravitation, the questions concerning the internal structure of matter have not been touched upon. Mr.\ G.\ \so{Mie} has proceeded in a different manner, developing his theory of gravitation in a close relation to his theory of matter\footnote{\so{G. Mie}, Ann.\ d.\ Phys.\ \textbf{37}, p.\ 511, 1912; \textbf{39}, p.\ 1, 1912; \textbf{40}, p.\ 1, 1913.}. However, I will use this opportunity to point out, that the expression (\ref{EQ:3}) for the components of the five-vector $\mathfrak{s}$ may provide a starting point for developing a theory of matter. By multiplying these expressions with the denominator on the right they yield five equations

\begin{equation}
\label{EQ:27}
\left\{
\begin{aligned}
(\mathfrak{s}_x \varPhi_x + \mathsf{S}_{xx}) \mathfrak{n}_x +
(\mathfrak{s}_x \varPhi_y + \mathsf{S}_{xy}) \mathfrak{n}_y +
(\mathfrak{s}_x \varPhi_z + \mathsf{S}_{xz}) \mathfrak{n}_z +
\\
(\mathfrak{s}_x \varPhi_u + \mathsf{S}_{xu}) \mathfrak{n}_u +
(\mathfrak{s}_x \varPhi_w + \mathsf{S}_{xw}) \mathfrak{n}_w = 0, 
\\
\dots \dots
\\
(\mathfrak{s}_w \varPhi_x + \mathsf{S}_{wx}) \mathfrak{n}_x +
(\mathfrak{s}_w \varPhi_y + \mathsf{S}_{wy}) \mathfrak{n}_y +
(\mathfrak{s}_w \varPhi_z + \mathsf{S}_{wz}) \mathfrak{n}_z +
\\
(\mathfrak{s}_w \varPhi_u + \mathsf{S}_{wu}) \mathfrak{n}_u +
(\mathfrak{s}_w \varPhi_w + \mathsf{S}_{ww}) \mathfrak{n}_w = 0. 
\end{aligned}
\right.
\end{equation}

For a theory of matter it is close at hand to make the ansatz that the components of $\mathsf{S}$ can be expressed in terms of $\mathfrak{s}$ and $\varPhi$. Hereby one has to recognize the condition that the determinant of the system (\ref{EQ:27}) must vanish, since the components of $\mathfrak{n}$ cannot all be equal to zero. The simplest ansatz would be\footnote{The agreement of this ansatz with the theory of \so{Mie} is noteworthy, though there are many differences between the two theories.}:

\begin{equation}
\label{EQ:28}
\left\{
\begin{aligned} 
\mathsf{S}_{xx} &= - \mathfrak{s}_x \varPhi_x,
\\
\mathsf{S}_{xy} &= - \frac{1}{2} 
(\mathfrak{s}_x \varPhi_y +
\mathfrak{s}_y \varPhi_x) \quad \mbox{etc}.
\end{aligned}
\right.
\end{equation}

If the components of the five-vectors $\mathfrak{s}$ and $\varPhi$ are proportional to each other

\begin{equation}
\label{EQ:28a}
\tag{28 a}
\frac{\mathfrak{s}_x}{\varPhi_x} =
\frac{\mathfrak{s}_y}{\varPhi_y} =
\frac{\mathfrak{s}_z}{\varPhi_z} =
\frac{\mathfrak{s}_u}{\varPhi_u} =
\frac{\mathfrak{s}_w}{\varPhi_w},
\end{equation}

then the expressions (\ref{EQ:28}) satisfy the equations (\ref{EQ:27}) \so{identically}. Making the ansatz (\ref{EQ:28}) we have to assume the aforementioned proportionality, since otherwise it would follow from (\ref{EQ:27}) that $\mathfrak{s}$ as well as $\varPhi$ would be orthogonal to  
$\mathfrak{n}$, contrary to the reality.

We will content ourselves with these hints of the possibility of a theory of matter. I plan to return to this question in a future communication.

The five dimensional outlook suggests also new possibilities in other directions. We have assumed that the four dimensional world-surface is planar, and that that the derivatives of all quantities in the normal direction of the surface vanish. One might however imagine that these assumptions in reality are only approximately satisfied. When making modifications along such lines one has to recognize above all that the condition of causality in the world-surface remains valid. Anyway, there is a multitude of possible modifications of the theory of gravitation which satisfy all our five dimensional equations.

There is still another kind of a modification that I will mention. Of the components of the five-potential the components normal to the world-surface will in all practical cases largely dominate the other components, and if we use $\varPhi$ in place of $\mathfrak{n}$ in the expression (\ref{EQ:3}) it will change extremely little. One may thus instead of (\ref{EQ:3}) assume for the five-current the expression

\begin{equation*}
\mathfrak{s}_x = -
\frac{\mathsf{S}_{xx} \varPhi_x + \mathsf{S}_{xy} \varPhi_y + \mathsf{S}_{xy} \varPhi_y + \mathsf{S}_{xu} \varPhi_u + \mathsf{S}_{xw} \varPhi_w}{\varPhi_x^2 + \varPhi_y^2 + \varPhi_z^2 + \varPhi_u^2 + \varPhi_w^2},
\end{equation*}
  
etc., and thus obtain a modified theory of gravitation.

With the foregoing account I hope to have shown, that the five dimensional perspective, while still having only a formal content, may however provide footholds for further developments of the theory of gravitation, which otherwise would have been hard to find. 
\\[10pt] {\hspace*{20pt}\so{Helsingfors}, September 1914.}
 
\end{document}